\documentclass[sigconf, authorversion=true, nonacm=true]{acmart}

\usepackage{graphicx}
\usepackage{amsmath}
\usepackage{amsfonts}
\usepackage{algorithm}
\usepackage{algpseudocode}

\AtBeginDocument{%
  \providecommand\BibTeX{{%
    \normalfont B\kern-0.5em{\scshape i\kern-0.25em b}\kern-0.8em\TeX}}}
    
\setcopyright{none}

\acmYear{2023}

\received{09 Oct 2023}

\setcopyright{none}
\renewcommand\footnotetextcopyrightpermission[1]{} 
\begin{document}

\title{A Possibility Frontier Approach to Diverse Talent Selection}

\author{Kadeem Noray}
\authornote{Both authors contributed equally to this research.}
\email{knoray@g.harvard.edu}
\affiliation{
    \institution{National Bureau of Economic Research}
    \country{USA}
}
\affiliation{
    \institution{Harvard University}
    \country{USA}
}

\author{Neil Natarajan}
\authornotemark[1]
\email{neil.natarajan@new.ox.ac.uk}
\affiliation{
    \institution{University of Oxford}
    \country{UK}
}

\date{September 2024}

\begin{abstract}
Organizations (e.g., talent investment programs, schools, firms) are perennially interested in selecting cohorts of \emph{talented} people. And organizations are increasingly interested in selecting \emph{diverse} cohorts. Except in trivial cases, measuring the tradeoff between cohort diversity and talent is computationally difficult. Thus, organizations are presently unable to make Pareto-efficient decisions about these tradeoffs. We introduce an algorithm that approximates upper bounds on cohort talent and diversity. We call this object the selection possibility frontier (SPF). We then use the SPF to assess the efficiency of selection of a talent investment program. We show that, in the 2021 and 2022 cycles, the program selected cohorts of finalists that could have been better along both diversity and talent dimensions (i.e., considering only these dimensions as we subsequently calculated them, they are Pareto-inferior cohorts). But, when given access our approximation of the SPF in the 2023 cycle, the program adjusted decisions and selected a cohort on the SPF. 
\end{abstract}




\maketitle

\section{Introduction}\label{sec:introduction}

Organizations (e.g. schools, firms, etc.) are perennially interested in selecting \emph{talented} people. And, as discrimination declines \cite{hsieh2019allocation} and social preferences for inclusion grows \cite{minkin2023diversity}, organizations are increasingly interested in selecting diverse cohorts \cite{dwork2012fairness}. But, in computer science, it is well known that both selecting cohorts to maximize diversity and measuring the tradeoff between diversity and average cohort talent are formally complex problems \cite{nemhauser1978analysis, huppenkothen2020entrofy}. While work exists tackling the problem of measuring diversity in talent identification, little of it approaches the problem from a computational angle, despite evidence that algorithmic approaches to talent identification outperform alternative approaches by as much as $50\%$ \cite{kuncel2013mechanical,li2020hiring}. This paper fills this hole by developing an algorithm for approximating the diversity-talent frontier and presenting suggestive case-study evidence that organizations make more efficient selection decisions when they have access to this technology. 

We start by defining the decision problem organizations face and demonstrating that, except in trivial cases, measuring the selection tradeoff between a cohort-level diversity and talent is formally complex. This implies that organizations do not know the talent cost of increasing diversity, which may result in Pareto-inferior selection decisions (i.e. selected cohorts could be better on one dimension without reducing the other dimension). We then introduce an algorithm that approximates the maximum diversity conditional on average cohort talent, which we call the selection possibility frontier (SPF). Given an organization's true preferences over diversity and performance, the SPF can be used to quantify the extent of the inefficiency in previously chosen cohorts and to allow organizations to improve decisions. We demonstrate this using data from an anonymized talent investment program, where we show that in the program's 2021 and 2022 application cycles, finalist cohorts were selected that could have been more diverse without reducing average talent (and vice-versa). We also show that, in the program's 2023 cycle, when the program had access to the SPF, the program chose to select a finalist cohort on the frontier. 

To begin, we demonstrate that selecting for diverse cohorts is formally complex, and therefore costly to solve. In particular, we reproduce a result from \citet{huppenkothen2020entrofy} to show that selecting a cohort that most closely matches a set of target proportions over various non-mutually exclusive attributes is an NP-hard problem. We then extend this result to show that computing the maximum combination of diversity and average cohort performance (i.e. the SPF) is also NP-hard unless no weight is given to diversity. We then apply results from optimization theory to demonstrate that, as long as the diversity objective function is submodular, monotonic, and non-negative, a greedy-style algorithm can be used to approximate the maximum of this function \cite{huppenkothen2020entrofy}. We use this finding to develop an algorithm that performs simultaneous greedy optimization of cohort diversity and mean cohort performance, which yields an estimate of the SPF. We conclude the theoretical component of the paper by noting that the error in the approximated SPF can be formally bounded, which gives decision-makers a sense of how close to the true frontier their decisions are. 

We conclude the paper with an application of the SPF algorithm to cohort selection for a real, anonymous talent investment program. As part of the program's 2023 application cycle, we worked with the program to define SPF-compatible metrics and construct appropriate frontiers. We leverage these SPFs to show that, in the 2021 cycle of the program, the set of selected finalists could have been $18.5\%$ more diverse without any reduction in performance and $17.8\%$ higher-performing without any reduction in diversity; in the 2022 cycle of the program, the set of selected finalists could have been $14.6\%$ more diverse without any reduction in performance and $24.1\%$ higher-performing without any reduction in diversity. Thus, both cohorts of finalists chosen were Pareto inferior by the program's own subsequent definitions of diversity and performance\footnote{It should be noted here that the definitions used here were not the only ones the program used, nor are these two dimensions the only ones in consideration. In practice, the program considers other factors and makes decisions holistically about applicant inclusion.}. We then show that in the 2023 cycle, when the program was given access to the SPF, they eliminated the inefficiency, choosing a cohort that was on the SPF. This provides suggestive evidence that the SPF aided their decision-making and allowed them to optimize in the face of a complex choice.

This paper is closely related to work in computer science on machine assistance in talent selection \cite{dwivedi2021artificial, tambe2019artificial, raghavan2020mitigating}. These systems solve for a variety of problems, from the difficulty of pre-screening to limited human accuracy when considering many factors \cite{tambe2019artificial}; while humans fail to properly consider these group-level considerations, machines excel at this sort of set optimisation \cite{krause2014submodular}. Some of these machine systems even advertise that they account for fairness and diversity considerations, but generally they fail to properly account for the trade-offs between these desiderata and the other dimensions (e.g. talent, productivity, etc.) that organizations value \cite{raghavan2020mitigating, gillet2011diversity,huppenkothen2020entrofy}. This paper adds to this literature by developing an algorithm that can be used across various contexts to estimate the maximum diversity possible conditional on average cohort performance, which provides decision-makers access to an estimate of the cost of diversity in their context. 

This paper is also closely related to \citet{kleinberg2018algorithmic}, which introduces an algorithm designed to calculate the max achievable performance at any threshold of representation for a minority group. \citet{kleinberg2018algorithmic} present a simple algorithm: define a minority group (e.g., women and non-binary people) and the corresponding majority group (e.g., men), rank individuals by some performance metric within each group, set a target threshold for the proportion of those selected who should be part of the minority group, select minority group members until that threshold is met, and finally, select the highest performance majority group members until all slots are filled. The key assumption that allows this transparent procedure to work is the mutual exclusivity of the minority group and the majority group. If diversity preferences are instead over non-mutually exclusive characteristics (e.g., race and gender), this procedure no longer works. This paper extends \citet{kleinberg2018algorithmic} by outlining the conditions under which the diversity-performance tradeoff can be estimated regardless of the number or type of identity characteristics organizations seek diversity over.

This paper also contributes to work on how to make selection processes fair and equitable. Much of this work is rooted in the ``similar treatment principle'' whereby individuals who perform similarly on some dimension are treated equally in selection \cite{dwork2012fairness}. This approach has been criticized, however, on the grounds that in circumstances when performance and demographic factors are correlated, employing this principle could substantially reduce the diversity of the chosen cohort, which may not accord with a decision-makers fairness preferences \cite{fleisher2021s}. Furthermore, when measurements of performance are heterogeneously biased in favour of certain demographic groups, similar treatment on the basis of these measurements leads to applicants with germane similarities being treated differently \cite{fleisher2021s}. An alternative perspective is to assume fairness and equity are captured by diversity, and instead to measure diversity directly. Three measures are commonly considered: (1) the majority-minority approach, (2) the generalized variance approach, and (3) the entropy statistic (see \citet{budescu2012measure} for a summary). Following \citet{huppenkothen2020entrofy}, we measure diversity using an entropy-based approach. We do this for two reasons. First, as \citet{huppenkothen2020entrofy} demonstrate, versions of an entropic scoring function have mathematical properties (i.e. submodularity and monotonicity) that allow us to bound the error of our SPF estimates. Second, preferences for diversity often take the form of ``closeness to a proportional target for each relevant identity characteristic'', which can be represented using entropic functions. Thus, our work adds to the applications of entropy-based methods for measuring diversity.

This paper is also closely related to the growing literature in economics demonstrating that selection technologies can help organizations to select closer to the frontier of diversity and performance. An early paper on this topic is \citet{autor2008does}, which shows that adding personality tests to the job applicant screening process in a national retail firm increased the productivity of selected workers without reducing minority representation among hires. More recently, \citet{li2020hiring} demonstrate that,  within a Fortune 500 tech company, diversity and productivity of hires can be improved by using exploration-based applicant screening algorithms that recommend applicants with rare sets of characteristics at higher rates in order to improve noisy predictions and, therefore, learn how to find promising applicants from atypical backgrounds. Similarly, \citet{bergman2021seven} show that, across seven colleges, using a prediction algorithm instead of a test to decide whether to place students into remedial classes improves student performance and increases minority representation in college-level (non-remedial) courses. This paper adds to this work by developing an algorithm that allows organizations to quantify how well they can possibly meet their diversity preferences and at what cost to performance. 

The remainder of the paper is organized as follows. Section \ref{sec:spf_def} defines the selection possibility frontier, including the diversity function and talent function we will use in our specific application. In Section \ref{sec:spf_alg}, we demonstrate the complexity of the cohort selection problem, formally present the algorithm that generates an estimate of the SPF, and formally derive bounds on our estimates. Section \ref{sec:case} applies the SPF procedure to characterize inefficiencies in selection in the program and subsequent improvements after using the SPF. In Section \ref{sec:disc}, we discuss our findings and reach a conclusion. 

\section{The Selection Possibility Frontier}\label{sec:spf_def}
We cannot meaningfully speak of an individual as diverse. Instead, diversity applies to groups of people. In the talent identification context, where we select a ``cohort'' of individuals from an ``applicant pool'', we are interested in both cohort-level diversity and pool-level diversity. Note that this is unlike applicant performance, which is primarily an individual-level property. However, in order to visualise the trade-offs between applicant performance and cohort diversity, we will need measurements of both either for candidates (creating candidate-level visualisations) or possible cohorts (creating cohort-level visualisations). As the ``diversity of a candidate'' is meaningless, we opt for a cohort-level performance metric.

While we might speak of the performance of a cohort, we usually mean this in aggregate. An individual will receive a performance score for a given metric, and group performance is a kind of average of individual performance.

\begin{figure*}[htb]
    \centering
    \includegraphics[width=0.6\textwidth]{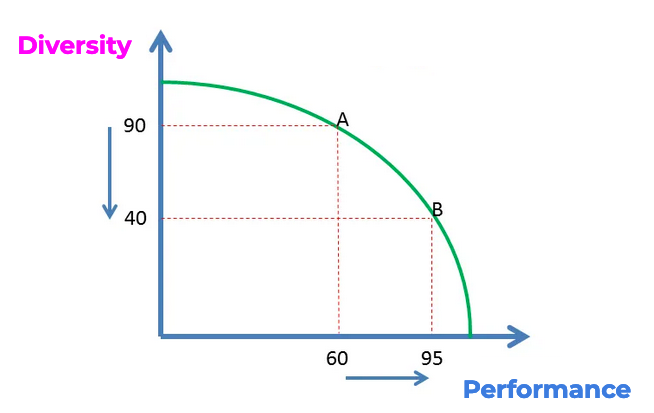}
    \caption{Example Selection Possibility Frontier}
    \label{fig:example_spf}
\end{figure*}

Informally, a Selection Possibility Frontier is a frontier curve relating the performance and diversity scores of possible cohorts. An example frontier is shown in figure \ref{fig:example_spf}.

\subsection{Formal Definitions}
Given a pool of candidates $U$, a `diversity' function $D$ with domain $\mathcal{P} (U)$ and finite, non-negative co-domain, and a `performance' function $P$ with the same domain and co-domain, a Selection Possibility Frontier (SPF) is a set of points $(p, d)$ that form a frontier relating the performance and diversity of each possible Pareto-efficient cohort $c$ in the cohort space $C$. More formally, a ``true'' $SPF$ is a set of ordered pairs meeting the conditions in equation \ref{eq:spf_defn}.

\begin{equation}
    \label{eq:spf_defn}
    \begin{split}
        &\forall (p, d) \in SPF \exists c \in C :(p, d) = \bigl( P(c), D(c) \bigr)  \\
        & \forall c \in C \forall (p, d) \in SPF :p \geq P(c) \lor d \geq D(c)
    \end{split}
\end{equation}

Note that the SPF, as defined above, contains all and only the Pareto-efficient cohorts. We note that these Pareto-efficient points are difficult to find (we show below that, for non-trivial diversity functions, finding these points is NP-hard). Thus, the true frontier is often approximated by an non-Pareto-efficient $SPF-approximation$. An $SPF-approximation$ is a set of ordered pairs meeting the conditions in equation \ref{eq:spf_approx_defn}. In practice, we often refer to $SPF-approximation$s simply as $SPF$s.

\begin{equation}
    \label{eq:spf_approx_defn}
    \begin{split}
        &\forall (p, d) \in SPF \exists c \in C :(p, d) = \bigl( P(c), D(c) \bigr)  \\
        &\forall (p, d) \in SPF \forall (p', d') \in SPF :p > p' \lor d > d'
    \end{split}
\end{equation}

When represented graphically, we may connect adjacent points on the $SPF$ to yield a curve like the one in figure \ref{fig:example_spf}. The approximation ratio of an $SPF-approximation$ is the minimum of the approximation ratios of the true $SPF$ to the points in the $SPF-approximation$.

\subsection{An Example}
We will soon introduce constraints on diversity, but for now, we might imagine a simple diversity function that measures gender parity, $D(c) = 1 - max(m(c) - nm(c), 0)$, where $m(c)$ and $nm(c)$ are the proportions of the chosen cohort $c$ that are male and non-male (grouping non-binary and female applicants) respectively. This function yields diversity $1$ when there are at least as many minority-gender applicants as male applicants, and yields less one point for each point of difference in proportion.

Similarly, simplest and most common such metric is the mean. For now, we imagine that this mean is the mean of some pre-existing performance metric that ranges from $[0,1]$.

Now that we have two cohort-level metrics that range from 0 to 1, we can visualise trade-offs between them by plotting possible cohorts on a graph and viewing the frontier. Note that while it is possible that the highest-scoring cohort also achieves gender parity, this is unlikely so long as our performance and diversity metrics are sufficiently demanding and discriminating. (In particular, if the performance metric is bijective, there will be only one top-performing cohort, and this cohort is unlikely to be optimally diverse).

\section{Calculating SPFs}\label{sec:spf_alg}
Ordinarily, we are interested in the maximal frontier of a given size $k$. I.e., $C$ is the set of subsets of $U$ of size $k$. Unfortunately, even under this condition, we cannot calculate the $SPF$ outright. Indeed, we demonstrate that finding any particular point on the frontier (save, under ordinary conditions, the highest-performance cohort) is NP-hard. To see why, consider the possible cohorts of size $k$ drawn from a pool of $n$ applicants. For even ordinary values of $n$ and $k$, this yields an absurd number of possible cohorts. (E.g., if selecting $2000$ people from a pool of $10000$, there are roughly $10^{2171}$ possible cohorts.) However, diversity is a property of cohorts. Thus, to actually find the most diverse cohort, we would require $10^{2171}$ calculations of diversity.

\subsection{A Proof of NP-Hardness}
Suppose we have an applicant pool $U$ of size $n$, i.e., $U = {a_1, a_2, ..., a_n}$. We can define a `diversity' function $D$ and a `performance' function $P$, both with domain $\mathcal{P} (U)$ and some non-negative, finite co-domain (for simplicity, we scale this to $[0, 1]$). Observe that any point on the frontier will be the maximum value of $F = \alpha D + (1 - \alpha) P$, for some $\alpha \in [0, 1]$.

Suppose that we desire an algorithm to find some cohort of size $k < n$ at a specific point on the frontier. Informally, as there are $n \choose k$ such cohorts, and we have no guarantees on the nature of $F$, we must call $F$ at least $n \choose k$ times, yielding an algorithm $\Omega(n^k)$.

More formally, if a function is an arbitrary function on sets, the cohort of size $k$ at the specified point on the frontier is the set $c$ that satisfies equation \ref{eq:diverse_set}. 

\begin{equation}
    \label{eq:diverse_set}
    \begin{split}
        \forall c' \in C : \bigl( |c'| = k \rightarrow  F(c') \leq F(c) \bigr)
    \end{split}
\end{equation}

Suppose further that there is an algorithm $A$ that purports to find $S$ with fewer than $n \choose k$ calls of $F$. Without loss of generality, suppose that $c = \{ a_1, ..., a_k \}$ is among the cohorts that is not an input to $F$. Crucially, $A$ must do this for any function $F$. Thus, let define particular diversity functions $F_1$ and $F_2$ in equation \ref{eq:d}. That is, for sets of size $k$, $D_1 = k$ specifically when $c = \{ a_1, ..., a_k \}$, and $F_1 = 0$ otherwise.

\begin{equation}
    \label{eq:d}
    \begin{split}
        F_1(c) &=
        \left\{
            \begin{array}{lr}
                k, & \text{if } c = \{ a_1, ..., a_k \}\\
                0, & \text{otherwise }
            \end{array}
        \right\} \\
        F_2(c) &=
        \left\{
            \begin{array}{lr}
                -k, & \text{if } c = \{ a_1, ..., a_k \}\\
                0, & \text{otherwise }
            \end{array}
        \right\}
    \end{split}
\end{equation}

As $A$ has not examined $F(\{ a_1, ..., a_k \})$, we cannot evaluate $F=F_1$ or $F=F_2$, and thus cannot return the correct answer. Note further that, as $\forall n : 0*n < k$, $A$ does not even approximate points on the frontier and any frontier constructed with $A$ is not even a guaranteed $n-approximation$.

\subsection{Approximating Cohort-Level Scores}

In order to achieve an approximation, we introduce two constraints on the frontier function: \textit{submodularity} and \textit{monotonicity}. Informally, these jointly guarantee that adding an applicant to a group never decreases its diversity, and always increases it by at least as much as adding that same applicant to a larger second group containing the first. More formally, submodularity and monotonicity are defined in equations \ref{eq:submodularity} and \ref{eq:mononicity} respectively.

\begin{equation}
    \label{eq:submodularity}
    \forall Y \subseteq U, X \subseteq Y, x \in U \setminus Y: D(X \cup \{x\}) - D(X) \leq D(Y \cup \{x\}) - D(Y)
\end{equation}

\begin{equation}
    \label{eq:mononicity}
    \forall Y \subseteq U, X \subseteq Y: D(X)\leq D(Y)
\end{equation}

Note that both of these conditions are preserved under function addition. It is thus trivial to show that when both $D$ and $P$ are submodular and monotone, $F$ is as well. We focus on applying these conditions to $D$ and $P$.

\subsubsection{A Formula for Diversity Functions}
This constraint on our diversity function means that our toy function, no longer suffices. Indeed, any sensible diversity function based on proportions will not suffice, as a larger cohort might be less proportionally diverse.

However, typically, organizations aim to select cohorts of a specific size. Thus, we need not consider all cohorts, and can instead consider cohorts of size $k$. On a cohort of size $k$, some proportional condition, e.g., $c_{prop} > x$ is identical to the condition $c_{count} > x*k$. We also note that, when a set of conditions forms a partition (e.g. applicants that are male and applicants that are not male), their proportions must always sum to $1$.

\begin{equation}
    \label{eq:prop}
    \begin{split}
        D(c) &= 1 - max(m_{prop}(c) - nm_{prop}(c), 0) \\
        &= 1 - max(1 - 2*nm_{prop}(c), 0) \\
        &= max(2*nm_{prop}(c), 1) \\
        &= \frac{max(2*nm_{count}(c), k)}{k}
    \end{split}
\end{equation}

With this in mind, the example function from above can be replaced with a count-based condition in equation \ref{eq:prop}. We note here that this function is both submodular and monotone. Furthermore, this process can be applied to \textit{any} proportional target. In general, if we desire that attribute $att$ have proportion $p$, we can set the target:

\begin{equation}
    \begin{split}
        D_{att}(c) &= max(att_{prop}(c), p) \\
        &= \frac{max(att_{count}(c), p*k)}{k}
    \end{split}
\end{equation}

Furthermore, as monotonicity and submodularity are both preserved under function addition, we can add targets together to get a larger diversity target, then scale that target so that it ranges across appropriate values.

We should also note that we might wish to have a target that is \textit{non-proportional} in nature. We present one such type of target, though interested organizations may, in theory, set any such target. We consider the case where a program wishes to select applicants from at least $m$ different countries, where Boolean function $in_n(a)$ is true if and only if applicant $a$ is in country $n$. Let $N$ be the set of all countries. Then our target is:

\begin{equation}
    \begin{split}
        D_N(c) &= max(\sum_{n \in N} \bigvee_{a \in c} in_n(a), m)
    \end{split}
\end{equation}

We note that this is already monotone and submodular.

\subsubsection{Compatible Performance Functions}
We note that, so long as it yields interval data, any individual-level scoring function suffices for our purposes. However, that does not mean that any \textit{aggregate} function of these individual-level scores will suffice. Indeed, the arithmetic mean of individual-level scores, which we use in the above example, is neither submodular nor monotone, and thus cannot be used in our search function.

This does not, however, necessitate major changes to our approach. Again, we ordinarily only consider cohorts of size $k$. When restricted to cohorts of size $k$, the aggregation functions $sum()$ and $mean()$ yield the same ordering, differing only by a constant factor of $k$. Furthermore, $sum()$ is both submodular and monotone. Thus, if we wish to optimise for the mean score, we may use $sum()$ as our accumulator instead.

\subsubsection{The Greedy Method}
Supposing, now, that we have two submodular and monotone functions $D$ and $P$, we present a greedy algorithm for $\bigl( 1-\frac{1}{e} \bigr)$-approximating a frontier (algorithm \ref{alg:frontier}).

\begin{algorithm}
    \caption{Greedy Frontier Optimisation}\label{alg:frontier}
    \begin{algorithmic}
        \Require $n \geq 0$, $D$, $P$, $U$
        \State $Cs \gets []$
        \For{$i \gets [0, n]$}
            \State $F_i \gets \frac{i*P}{n}+\frac{(n-i)*D}{n}$
            \While{$|C| < n$}
                \State $C \gets C \cup \Bigl\{ \texttt{argmax}_{i \in U \setminus C}\bigl(F_i(C \cup \{i\} ) \bigr) \Bigr\}$
            \EndWhile
            \State $Cs.\texttt{append}(C)$
        \EndFor
    \end{algorithmic}
\end{algorithm}

Crucially, each $F_i$ is submodular and monotonic, as these properties are closed under function addition and positive scalar multiplication. It is well-known that the greedy algorithm yields a $\bigl( 1-\frac{1}{e} \bigr)$-approximation of any submodular, monotonic set function \cite{bordeaux_submodular_2014}. That is, the algorithm selects cohorts whose $F_i$ values are at least $\frac{1}{1-\frac{1}{e}}$ of the maximum $F_i$ any cohort of that size selected from the same applicant pool. Thus, the Greedy Frontier Optimisation algorithm returns points on a curve that $\bigl( 1-\frac{1}{e} \bigr)$-approximates the true SPF\footnote{In practice, the outputs of the greedy algorithm do not always themselves form a convex curve. We remove produced points that do not sit on the convex curve.}. We note that this is a worst-case approximation ratio, and that the actual approximation ratio may be much better.

\section{A Case Study With A Talent Investment Program}\label{sec:case}

Now we turn to demonstrating how the SPF algorithm can be used to better understand past selection decisions and to improve talent selection choices. To do this, we leverage data from a talent investment program. As this program wishes to remain anonymous, we refer to it simply as ``the program''. The program seeks to find and support talented young people.

In our analysis, we use data from the program's 2021, 2022, and 2023 application cycles. In particular, we leverage data on finalist selection, which is when the pool of $2000$ to $3000$ applications previously selected for second-round review is further reduced to $500$ finalists. This selection stage is informed by psychometric assessments and application reviews from multiple reviewers. The data we have access to consists of the performance scores of applicants as well as relevant demographic characteristics (e.g., gender identity, nationality, parental education, socioeconomic status, etc.). 

With these data, we can estimate the SPF for finalist selection for each application cycle. Though an organizations selection preferences are not generally available, in this case we had
access to the program's stated target for gender, country, and socioeconomic status balance\footnote{To maintain the anonymity of the program and the integrity of their selection process, the exact details of these diversity targets are excluded from the public-facing version of this paper. Additional detail may be shared in consultation with the program upon request.}. This allows us to compute an entropic diversity function that corresponds to their diversity preferences. And, following the program's actual finalist selection criteria, we take the mean of individual-level performance scores as our proxy to construct our cohort-level talent scores. 

In our analysis, we exploit a novel feature of this case: between finalist selection of the Cycle 2022 and Cycle 2023 cohorts, this program's selection team was given access to an estimate of the SPF. Thus, in our analysis, we expect to see finalist cohorts selected well within the SPF for the first two cycles, but marked improvement in the efficiency of selection in the third cycle. This allows us to showcase the SPF as both a diagnostic tool that quantify the inefficiency of past selection decisions as well as a tool for improving those decisions. 

\begin{figure}[htb]
    \centering
    \includegraphics[width=.4\textwidth,keepaspectratio]{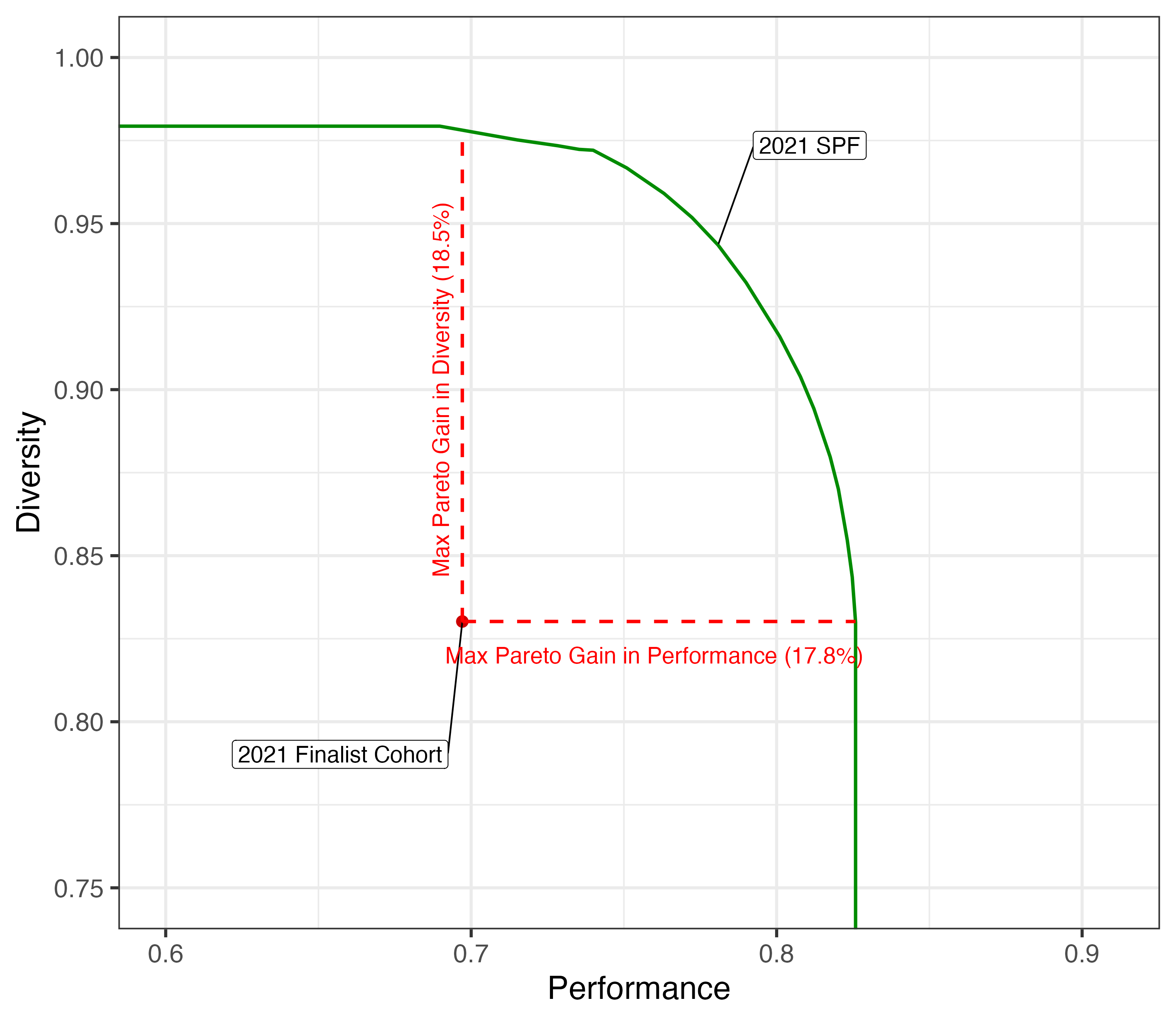} 
    \caption{Cycle 2021 Finalist Selection SPF}\label{fig:yr1_spf}
\end{figure}

Figure \ref{fig:yr1_spf} displays the SPF we estimate for the Cycle 2021 finalist cohort. The y-axis represents the diversity score while the x-axis represents mean cohort performance. The green curve is our estimate of the Cycle 2021 SPF, which represents the upper bound of diversity that is achievable at every level of cohort performance. The red dot depicts the actual level of diversity and performance of the finalists that were selected in Cycle 2021; clearly, the cohort selected was within the frontier. As a way of quantifying the inefficiency, the vertical and horizontal dashed red lines represent the maximum Pareto gain that was possible along the diversity and performance dimensions respectively. In particular, using our diversity and performance metrics, cohort diversity could have been improved by $18.5\%$ without any reduction in cohort performance. And, cohort performance could have been improved by $17.8\%$ without any cost to diversity. 

\begin{figure}[htb]
    \centering
    \includegraphics[width=.4\textwidth,keepaspectratio]{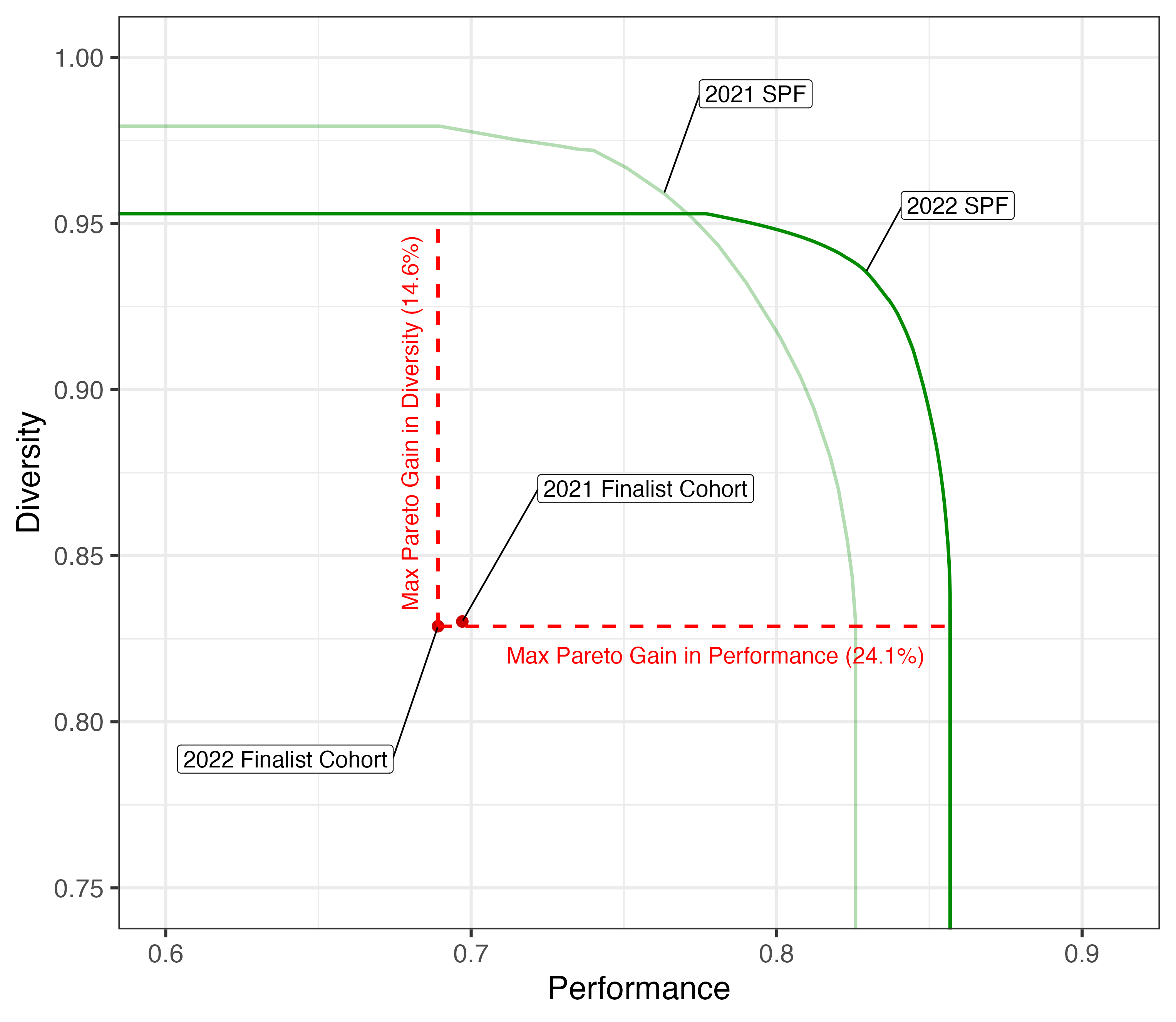}
    \caption{Cycle 2022 Finalist Selection SPF}\label{fig:yr2_spf}
\end{figure}

Figure \ref{fig:yr2_spf} adds the Cycle 2022 cohort SPF estimate and the Cycle 2022 finalist cohort levels of diversity and performance. In this application cycle, we see qualitatively similar results. In this case the max Pareto gain in diversity is 14.6\% while the max Pareto gain in cohort performance is 24.1\%.

\begin{figure}[htb]
    \centering
    \includegraphics[width=.4\textwidth,keepaspectratio]{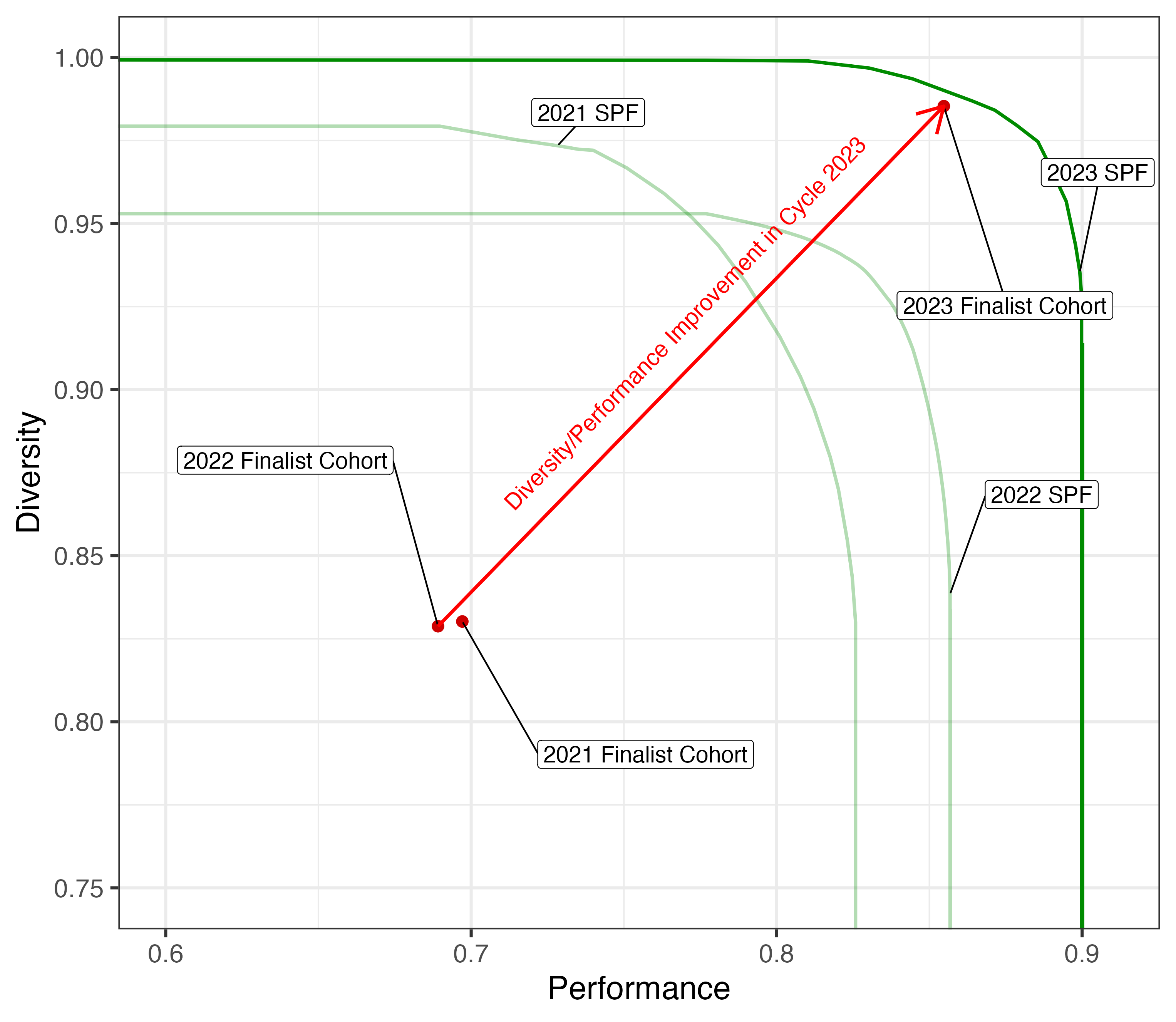}
    \caption{Cycle 2023 Finalist Selection SPF}\label{fig:yr3_spf}
\end{figure}

Finally, Figure \ref{fig:yr3_spf} adds the Cycle 2023 cohort SPF estimate and the Cycle 2023 finalist cohort levels of diversity and performance. Here we see marked changes in the selection patterns. In particular, the Cycle 2023 finalist cohort is essentially on the SPF, making the possible Pareto improvements negligible. This suggests two things. First, it provides evidence that selection decisions in Years 1 and 2 were, considering only the two desiderata formalised here, inefficient; had the program only known about the possibility of making Pareto improvements across these stated preferences, they may have made different decisions\footnote{It is possible that the actually chosen cohorts better satisfy other hidden preferences. In this case, the program considered qualitative factors about applicants and made decisions holistically}. Second, it provides evidence that the SPF is not merely a diagnostic tool, but can actually influence the decisions of talent selection agents.

\section{Discussion}\label{sec:disc}
In section \ref{sec:spf_def} above, we introduce the Selection Possibility Frontier, a novel method for quantifying and visualising the tradeoffs selection procedures make between applicant performance and cohort diversity. Following this, we demonstrate that calculating the SPF outright is NP-hard (as is calculating the most diverse cohort) and provide a means of approximating the SPF in section \ref{sec:spf_alg}. Finally, in section \ref{sec:case}, we demonstrate the value of such an SPF in practice through a case study.

Our results indicate that the SPF method allows organizations to satisfy their preferences on both performance and diversity simultaneously and facilitates decision-making that better optimises for these two quantities.

\subsection{Future Work}
The greedy method yields a good approximation of our curve in polynomial time. And, indeed, it has among the best guarantees of worst-case approximation of any method. However, other methods such as Monte Carlo simulations also yield good approximations, and it remains to be seen if the greedy method tends towards the best performance in practice. In future we intend to test a number of optimisation procedures on real data to determine which method yields the most efficient and most accurate approximations.

Furthermore, though we have evidence that the SPF alters selection decision-making in such a way as to improve both performance and diversity, we do not yet have a thorough understanding of how or why. Furthermore, it is not clear how many of an organization's preferences are well-captured by our performance-diversity framing. Future work is needed to better understand how selection experts make these trade-offs and design tools around that knowledge.

\subsection{Conclusion}
While open questions exist about how and why the SPF alters decision-making, it is clear that the SPF is not merely a diagnostic tool, but can actually influence the decisions of talent selection agents. Furthermore, this influence appears to drive decisions towards Pareto-optimality aggregate applicant performance and cohort diversity. The Selection Possibility Frontier is a valuable tool for decision-makers and serves to advance research into organizational considerations of diversity in talent selection.

\clearpage
\bibliographystyle{ACM-Reference-Format}
\bibliography{main}

\end{document}